\documentclass{ws-procs975x65}
\usepackage{times}

\begin{document}

\title{Genotype-based Case-Control Analysis, Violation of
Hardy-Weinberg Equilibrium, and Phase Diagrams 
}

\author{Young Ju Suh 
}

\address{BK21 Research Division of Medicine and Department of 
Preventive Medicine, College\\
of Medicine, Ewha Womans University, Seoul, Korea.  E-mail: ysprite@hotmail.com }

\author{Wentian Li}

\address{The Robert S. Boas Center for Genomics and Human Genetics, Feinstein Institute\\
for Medical Research, North Shore LIJ Health System, 
Manhasset, NY 11030, USA\\ 
E-mail: wli@nslij-genetics.org}

\maketitle

\abstracts{
We study in detail a particular statistical method in 
genetic case-control analysis, labeled ``genotype-based 
association", in which the two test results from assuming 
dominant and recessive model are combined in one
optimal output. This method differs both from
the allele-based association which artificially doubles
the sample size, and the direct $\chi^2$ test on 3-by-2
contingency table which may overestimate the degree of freedom.
We conclude that the comparative advantage (or disadvantage)
of the genotype-based test over the allele-based test
mainly depends on two parameters, the allele frequency
difference $\delta$ and the Hardy-Weinberg disequilibrium
coefficient difference $\delta_\epsilon$. Six different
situations, called ``phases", characterized by the two $X^2$
test statistics in allele-based and genotype-based test, 
are well separated in the phase diagram parameterized by 
$\delta$ and $\delta_\epsilon$. For two major groups of
phases, a single parameter $\theta = tan^{-1} (\delta/\delta_\epsilon)$
is able to achieves an almost perfect phase separation.
We also applied the analytic result to several types
of disease models. It is shown that for dominant and
additive models, genotype-based tests are favored over
allele-based tests.
}

\section{Introduction }\label{sec:1}

Genetic association analysis is a major tool in mapping human
disease genes\cite{risch96,clayton05,wliweb}. A simple association
study is the case-control analysis, in which individuals with
and without disease are collected (roughly the equal number
of sample per group for an optimal design), DNA samples extracted
and genetic markers typed. The prototype of a genetic marker
is the two-allele single-nucleotide-polymorphism (SNP)\cite{brookes}.
If the two alleles are $A$ and $a$, there three possible genotypes:
$AA$, $Aa$, $aa$, consisting of the maternally-derived
and paternally-derived copy of an allele. The three genotype
frequencies are calculated in case (disease) and control (normal)
group, and a strong contrast of the two sets of genotype
frequencies can be used to indicate an association between
that marker and the disease. 

The statistical analysis in an association study seems to
be simple -- mostly the standard Pearson's $\chi^2$ test in categorical
analysis\cite{agresti}, there are nevertheless subtle differences 
among various approaches. Some people use the $2 \times 3$ genotype
count table to carry about test with $\chi^2$ distribution 
of $df=2$ degrees
of freedom\cite{burton}. This method may overestimate
the degree of freedom if the Hardy-Weinberg equilibrium holds
true. Other people use the allele-based test, where each
person contributes two allele counts, and the allele frequency
is compared in a $2 \times 2$ allele count table.  This approach
artificially doubles the sample sizes without a theoretical
justification\cite{sasieni}. A third approach, what we called
``genotype-based" case-control association analysis, remains faithful to 
the sample size, while does not overestimate the degrees of freedom. 

A genotype-based analysis can be simply summarized here.
Two Pearson's $\chi^2$ tests are carried out on two $2 \times 2$ count
tables: the first is constructed by combining the $AA$ and $Aa$
genotype counts and keeping the $aa$ genotype column, and
the second by combining the $Aa$ and $aa$ genotype counts.
If the marker happens to be the disease gene and $A$ is the
mutant allele ($a$ is the wild type allele), then the first
table is consistent with a dominant disease model, whereas
second a recessive disease model. The two $\chi^2$ tests
lead to two $p$-values, and the smallest one (the more
significant one) is chosen as the final test result.

Genotype-based analysis has been used in practice 
many times\cite{yamada,tokuhiro,lee}, without a particular
name, and without a theoretical study. In this article,
we will take a deeper look of the genotype-based analysis.
We will show that the justification of using genotype-based
tests is intrinsically related to the Hardy-Weinberg
disequilibrium, but there are more than just a non-zero
Hardy-Weinberg disequilibrium coefficient that is important. 

The article is organized as follows: we first show that there is
no advantage in using genotype-based test if there is
no Hardy-Weinberg disequilibrium; we then examine the situation
with Hardy-Weinberg disequilibrium, and use the
two parameters, the allele frequency difference and
the difference of two Hardy-Weinberg disequilibrium
coefficients, to construct a phase diagram; the phase
diagram is further simplified by using just one parameter;
our analytic result is illustrated by a real example
from the study of rheumatoid arthritis; we apply the
formula to different models; and finally future works
are discussed.

\vspace{-0.1in}

\section{No advantage for genotype-based analysis if
Hardy-Weinberg equilibrium holds true exactly}\label{sec:2}

In an ideal situation, we assume $N$ case samples and
$N$ control samples, and the $A$ allele frequency in
case and control groups is $p_1$ and $p_2$ ($q_1=1-p_1, q_2=1-p_2$).
On average (or in the asymptotic limit), the allele
and genotype counts are listed in Table \ref{table:1} where
the Hardy-Weinberg equilibrium (HWE) is assumed.

For a $\{ N_{ij} \}$ ($i,j=1,2$) 2-by-2 contingency table, the
Pearson's (O-E)$^2$/E (O for observed count, and E for
expected count) test statistic is: 
\begin{equation}
\label{eq:def-x2}
X^2= \frac{ (N_{11}N_{22}-N_{12}N_{21})^2(N_{11}+N_{12}+N_{21}+N_{22})}
 { (N_{11}+N_{12})(N_{21}+N_{22})(N_{11}+N_{21})(N_{12}+N_{22})}
\end{equation}
Using the table elements in Table \ref{table:1}, we can derive
\begin{eqnarray}
\label{eq:x2}
X^2_{\rm allele} &=& \frac{ (2N)^4 (p_1q_2-p_2q_1)^2 4N}
{ (2N)^4 (p_1+p_2)(q_1+q_2)}
 = \frac{ 4N (p_1-p_2)^2}{ (p_1+p_2)(q_1+q_2) }
 \nonumber \\
X^2_{\rm dom} &=&
 \frac{N^4 [ (p_1^2+2p_1q_1)q_2^2 - (p_2^2+2p_2q_2)q_1^2]^2 2N}
{ N^4 (p_1^2+2p_1q_1+p_2^2+2p_2q_2)(q_1^2+q_2^2)}
= \frac{2N (p_1-p_2)^2(q_1+q_2)^2}{ (2-q_1^2-q_2^2)(q_1^2 + q_2^2)}
 \nonumber \\
X^2_{\rm rec} &=&
 \frac{N^4 [ (q_1^2+2p_1q_1)p_2^2 - (q_2^2+2p_2q_2)p_1^2]^2 2N}
{ N^4 (q_1^2+2p_1q_1+q_2^2+2p_2q_2)(p_1^2+p_2^2)}
= \frac{2N (q_1-q_2)^2(p_1+p_2)^2}{ (2-p_1^2-p_2^2)(p_1^2 + p_2^2)}
\end{eqnarray}
To further simplify the notation, let's denote $\delta \equiv p_1 -p_2$
as the allele frequency difference, $\overline{p} \equiv (p_1+p_2)/2$
as the averaged $A$ allele frequency across groups, 
and the averages of the squared terms $\overline{p^2} \equiv (p_1^2+p_2^2)/2$
($\overline{q}$ and $\overline{q^2}$ are defined similarly).
Then Eq.(\ref{eq:x2}) becomes: 
\begin{eqnarray}
X^2_{\rm allele} &=& \frac{N \delta^2}{ \overline{p} \cdot \overline{q}},
 \nonumber \\
X^2_{\rm dom} &=& \frac{2N \delta^2 \overline{q}^2}{\overline{q^2} (1- \overline{q^2} )},
 \nonumber \\
X^2_{\rm rec} &=& \frac{2N \delta^2 \overline{p}^2}{\overline{p^2} (1- \overline{p^2} )}.
\end{eqnarray}
Since the genotype-based test is determined by the maximum
value among $X^2_{\rm dom}$ and $X^2_{\rm rec}$, we would like
to prove an inequality between $X^2_{\rm allele}$ and max($X^2_{\rm dom}$,$X^2_{\rm rec}$). 

\begin{table}[bt]
\tbl{Count tables for genotype-based analysis under HWE
\label{table:1} }
{\begin{tabular}{@{}c|cc|cc|cc@{}}
\hline
{} & $A$ & $a$ & $AA+Aa$ & $aa$ & $AA$ & $AA+Aa$ \\[1ex]
{} & \multicolumn{2}{c|}{ allele count} & \multicolumn{2}{c|}{ dominant model}
 & \multicolumn{2}{c}{ recessive model} \\[1ex]
\hline
case & 2 $N p_1$ &  2 $Nq_1$ &  $N(p_1^2+2p_1q_1)$ & $N q_1^2$
 & N$p_1^2$ & N($2p_1q_1+q_1^2$) \\[1ex] 
control & 2 $N p_2$ &  2 $N q_2$ &  $N(p_2^2+2p_2q_2)$ & $N q_2^2$
 & $ N p_2^2$ & $ N(2p_2q_2+q_2^2)$ \\[1ex]
\hline
\end{tabular}
}
\end{table}

Towards this aim, we first compare $X^2_{\rm allele}$ and $X^2_{\rm dom}$. Due to
the following two inequalities:
\begin{eqnarray}
\overline{q^2} &=&\frac{2q_1^2+2 q_2^2 }{4}
\ge \frac{2q_1^2+2 q_2^2 - (q_1-q_2)^2 }{4} = \frac{ (q_1+q_2)^2}{4}
= \overline{q}^2
\nonumber \\
2 \overline{p} \cdot \overline{q} &=&  2 \overline{q} -  2 \overline{q}^2
= q_1 +q_2 -  (q_1q_2 + \overline{q^2})
= 1- \overline{q^2} - (1-q_1)(1-q_2) \le
1- \overline{q^2}, \nonumber
\end{eqnarray}
we have
\begin{equation}
\frac{\overline{q^2}}{ 2 \overline{p} \cdot \overline{q}} \ge \frac{ \overline{q}^2}{ 1- \overline{q^2}},
\end{equation}
which leads to $X^2_{\rm allele} \ge X^2_{\rm dom}$.
The similar approach shows that $\overline{p^2} \ge \overline{p}^2$
and $2 \overline{p} \cdot \overline{q}  \le 1- \overline{p^2}$,
which leads to $X^2_{\rm allele} \ge X^2_{\rm rec}$.

With the proof that $X^2_{\rm allele} \ge \max(X^2_{\rm dom}, X^2_{\rm rec})$, we have
shown that allele-based $X^2$ ($p$-value) is always larger 
(smaller) than the genotype-based $X^2$ ($p$-value).
In other words, if HWE holds exactly
true, there is no need to carry out a genotype-based association
analysis. To certain extend, this result is not surprising
since allele-based test utlizes twice the number of samples
as the genotype-based test, even though the latter has
one advantage of testing multiple (two) disease models.
Clearly, the increase in sample size more than compensates 
the advantage of testing multiple models, when HWE is true.

\begin{table}[bt]
\tbl{Count tables for genotype-based analysis under HWD
\label{table:2} }
{\begin{tabular}{@{}c|cc|cc|cc@{}}
\hline
{} & $A$ & $a$ & $AA+Aa$ & $aa$ & $AA$ & $AA+Aa$ \\[1ex]
{} & \multicolumn{2}{c|}{ allele count} & \multicolumn{2}{c|}{ dominant model}
 & \multicolumn{2}{c}{ recessive model} \\[1ex]
\hline
case & 2 $N p_1$ &  2 $N q_1$ &  $N(p_1^2+2p_1q_1-\epsilon_1)$ & $N(q_1^2+\epsilon_1)$
 & N($p_1^2+\epsilon_1$) & N($2p_1q_1+q_1^2-\epsilon_1$) \\[1ex] 
control & 2 $N p_2$ &  2 $N q_2$ &  $N(p_2^2+2p_2q_2-\epsilon_2)$ & $N(q_2^2+\epsilon_2)$
 & $N(p_2^2+\epsilon_2)$ & $N(2p_2q_2+q_2^2-\epsilon_2)$ \\[1ex]
\hline
\end{tabular}
}
\end{table}

\section{Adding violation of Hardy-Weinberg equilibrium}

The result in the previous section actually does not
disapprove the genotype-based association, since
HWE in real data is often violated, even if it is
not significantly violated. To characterize a realistic genotype
count table, one more parameter besides the allele frequency
is needed: the Hardy-Weinberg disequilibrium coefficient (HWDc)\cite{weir}.
The HWDc $\epsilon$ is defined as\cite{weir} 
$\epsilon= p_{AA} - p_A^2 = p_{aa}- p_a^2 = -(p_{Aa}- 2 p_a p_A)/2
= p_{aa}p_{AA} - p_{Aa}^2/4$.
For case and control groups, two HWDc's are used $\epsilon_1$
and $\epsilon_2$. The three count tables under HWD are now 
parameterized in Table \ref{table:2}.

Applying the definition of $X^2$ in Eq.(\ref{eq:def-x2}) to the
count tables in Table \ref{table:2} (note that the allele counts
are not affected by HWD), we have
\begin{eqnarray}
\label{eq:x2-epsilon}
X^2_{\rm allele} &=&  \frac{ 4N (p_1-p_2)^2}{ (p_1+p_2)(q_1+q_2) }
 \nonumber \\
X^2_{\rm dom, HWD} &=&
 \frac{N^4 [ (p_1^2+2p_1q_1-\epsilon_1)(q_2^2+\epsilon_2) - 
(p_2^2+2p_2q_2-\epsilon_2)(q_1^2+\epsilon_1)]^2 2N}
{ N^4 (p_1^2+2p_1q_1+p_2^2+2p_2q_2-\epsilon_1-\epsilon_2)(q_1^2+q_2^2
+\epsilon_1+\epsilon_2)}
 \nonumber \\
& =& \frac{2N [(p_1-p_2)(q_1+q_2) -(\epsilon_1-\epsilon_2)]^2}
 { (2-q_1^2-q_2^2 - \epsilon_1-\epsilon_2)(q_1^2 + q_2^2+ \epsilon_1+\epsilon_2)}
 \nonumber \\
X^2_{\rm rec, HWD} &=&
 \frac{N^4 [ (q_1^2+2p_1q_1 -\epsilon_1)(p_2^2+\epsilon_2) - (q_2^2+2p_2q_2-\epsilon_2)(p_1^2+\epsilon_1)]^2 2N}
{ N^4 (q_1^2+2p_1q_1+q_2^2+2p_2q_2 - \epsilon_1-\epsilon_2)(p_1^2+p_2^2 + \epsilon_1+\epsilon_2)}
 \nonumber \\
&=& \frac{2N [(q_1-q_2)(p_1+p_2)- (\epsilon_1-\epsilon_2)]^2}
{ (2-p_1^2-p_2^2 -\epsilon_1-\epsilon_2)(p_1^2 + p_2^2 +\epsilon_1+\epsilon_2)}
\end{eqnarray}
Again shorthand notations are introduced: $\delta_{\epsilon} \equiv \epsilon_1 - \epsilon_2$,
and $\overline{\epsilon} \equiv (\epsilon_1 + \epsilon_2)/2$. Eq.(\ref{eq:x2-epsilon})
is rewritten as
\begin{eqnarray}
\label{eq:x2-epsilon-2}
X^2_{\rm allele} &=& \frac{N \delta^2}{ \overline{p} \cdot \overline{q}},
 \nonumber \\
X^2_{\rm dom, HWD} &=&
\frac{2N  (\delta \overline{q} - \frac{ \delta_{\epsilon}}{2} )^2 }
 { (\overline{q^2} + \overline{\epsilon})
 (1- \overline{q^2} - \overline{\epsilon}) }
 \nonumber \\
X^2_{\rm rec, HWD} &=&
 \frac{2N (\delta \overline{p} - \frac{ \delta_{\epsilon}}{2}  )^2 }
 { (\overline{p^2} +\overline{\epsilon} )
 (1- \overline{p^2} - \overline{\epsilon}) }
\end{eqnarray}

From Eq.(\ref{eq:x2-epsilon-2}), it is not clear whether $X^2_{\rm allele}$ 
is still larger than $X^2_{\rm dom, HWD}$ and $X^2_{\rm rec, HWD}$. 
Systematic scanning of the 4-parameter space ($p_1, p_2, \epsilon_1, \epsilon_2$)
would offer a solution, but the result cannot be displayed
on a 2-dimensional space. In the following, we simplify the display
of the ``phase diagram" by using only two (or one) parameters.

\section{Phase diagram with one and two parameters}

The term ``phase diagram" is borrowed from the field of statistical
physics\cite{landau}. In a typical diagram used in statistical
or chemical physics, phases (e.g. solid, liquid and gas) as well
as phase boundaries (e.g. melting line) are displayed as a function
of physical quantities such as temperature and pressure. Phase transition occurs at phase
boundaries. For our topic, a phase indicates, for example,
whether allele-based or genotype-based test leads to a higher $X^2$ value;
or it can indicate whether or not the $X^2$ value leads to a statistically
significant result (e.g. $p$-value $<$ 0.05). 
The quantities chosen to mimic temperature or pressure for
our topic should highlight the phase separation and phase
transitions.

\begin{figure}[thp]
\epsfxsize=7cm   
 \centerline{\epsfxsize=3.8in\epsfbox{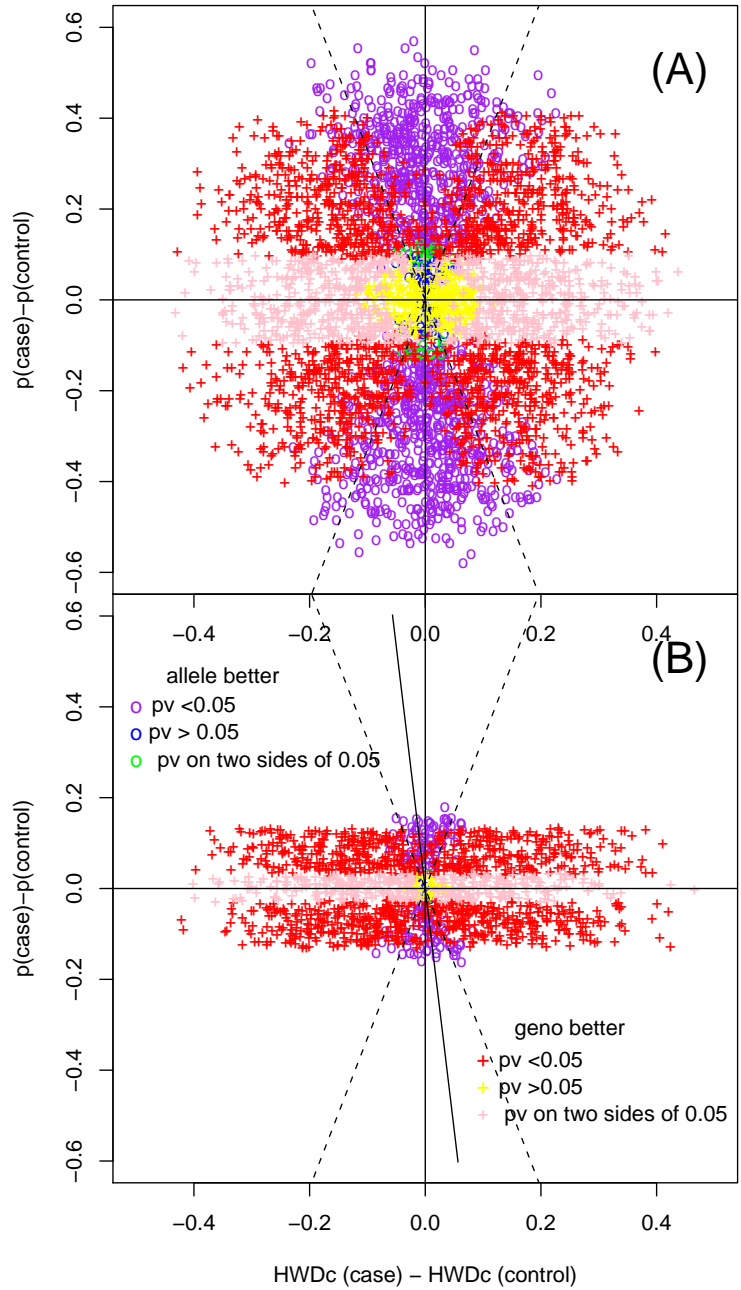}}
\caption{
The phase diagram parameterized by $\delta_\epsilon= \epsilon_1-\epsilon_2$ (x-axis)
and $\delta = p_1-p_2$ (y-axis), where $p$ is the allele frequency for $A$
and $\epsilon$ is the Hardy-Weinberg disequilibrium coefficient,
determined by a numerical simulation. (A) 100 samples per group with
5000 replicates (5000 points in the plot); (B) 1000 samples per group
with 5000 replicates.  Six phases are marked: 
I. $p$-value for genotype-based test is smaller than that for 
allele-based test (and both $p$-values are smaller than 0.05) (red cross);
II. similar to I, but both $p$-values are larger than 0.05 (yellow cross);
III. similar to I, but one $p$-value is smaller than 0.05 and another larger than 0.05
(pink cross);
IV. $p$-value for allele-based test is smaller than that
for genotype-based test (and both $p$-values are smaller than 0.05)
(purple circle);
V. similar to IV, but both $p$-values are larger than 0.05
(blue circle);
VI. similar to V, but one $p$-value is smaller than 0.05 and another larger than 0.05
(green circle). The two dashed lines have angle of
73.125$^\circ$ and $-$73.125$^\circ$, and the solid line has
angle of 95.37$^circ$. 
\label{fig:1}}
\end{figure}

Eq.(\ref{eq:x2-epsilon-2}) provides us a hint that the allele frequency
difference in two groups, $\delta$, and the HWDc difference,
$\delta_\epsilon$, could be good quantities for phase separation. First of all,
$\delta$ directly controls the magnitude of $X^2$, so it
should separate ``significant phases" from ``insignificant phases".
Secondly, the relative magnitude and sign of $\delta$ and $\delta_\epsilon$
seems to control the difference between $X^2_{\rm allele}$
and $X^2_{\rm dom, HWD}$ or  $X^2_{\rm rec, HWD}$, so it should
be a good quantity to separate ``favoring-allele-based-test phase"
(when $X^2_{\rm allele} > \max( X^2_{\rm dom, HWD}, X^2_{\rm rec,HWD})$)
and ``favoring-genotype-based-test phase"
(when $X^2_{\rm allele} < \max( X^2_{\rm dom, HWD}, X^2_{\rm rec,HWD})$).

We carried out the following simulation to construct the phase
diagram: 5000 replicates of case-control datasets with 100 cases
and 100 controls (in another simulation, the sample size is
1000 per group); For each replicate, the three genotypes are
randomly chosen, then the allele frequency and Hardy-Weinberg
disequilibrium coefficient were determined.  Fig.\ref{fig:1} shows the 
simulation result parameterized by $\delta_\epsilon$ 
(x-axis) and $\delta$ (y-axis). Six phases (labeled I-VI) are 
illustrated using 6 different colors, within the two larger categories: 
\begin{itemize}
\item
Favoring genotype-based tests (crosses in Fig.\ref{fig:1})
 \begin{itemize}
	\item I. $p$-values for both genotype- and allele-based tests are $<$ 0.05 (red)
	\item II. $p$-values for both genotype- and allele-based tests are $>$ 0.05 (yellow)
	\item III. $p$-value for genotype-based test is $<$ 0.05, that for
allele-based test is $>$ 0.05 (pink)
 \end{itemize}
\item
Favoring allele-based tests (circles in Fig.\ref{fig:1})
 \begin{itemize}
	\item IV. $p$-values for both genotype- and allele-based tests are $<$ 0.05 (purple)
	\item V. $p$-values for both genotype- and allele-based tests are $>$ 0.05 (blue)
	\item VI. $p$-value for allele-based test is $<$ 0.05, that for
genotype-based test is $>$ 0.05 (green)
 \end{itemize}
\end{itemize}
As can be seen from Fig.\ref{fig:1}, the two parameters, $\delta$ and
$\delta_\epsilon$ does a pretty good job in separating six different
phases, although minor overlap between phases occurs. The
overall performance of $\delta$ and $\delta_\epsilon$
as phase parameters is satisfactory.

As expected, the magnitude of $p$-values is mainly controlled by the
y-axis. Smaller allele frequency differences (smaller $\delta$'s)
result in non-significant $p$-values, and significant results
are located far away from the $\delta=0$ line. On the
other hand, the $\delta_\epsilon$ mainly controls whether 
allele-based or genotype-based test is more significant. 
However, $\delta_\epsilon$ itself is not enough: it acts 
jointly with $\delta$ to achieve the phase separation: for 
genotype-based test to have a smaller $p$-value than the allele-based test
and both are smaller than 0.05 (red points in Fig.\ref{fig:1}),
$\delta_\epsilon$ tends to have the different sign as that of
$\delta$.

The effect of sample size on the phase diagram can be
examined by comparing Fig.\ref{fig:1}(A) and Fig.\ref{fig:1}(B).
Phases II, III, V, VI all shrink in area simply because
a larger sample size is more likely to lead to
a $p$-value $<$ 0.05 replicate. The relative location
of different phases in Fig.\ref{fig:1} remains the same.

\begin{figure}[t]
  \begin{turn}{-90}
  \centerline{\epsfxsize=3.6in \epsfbox{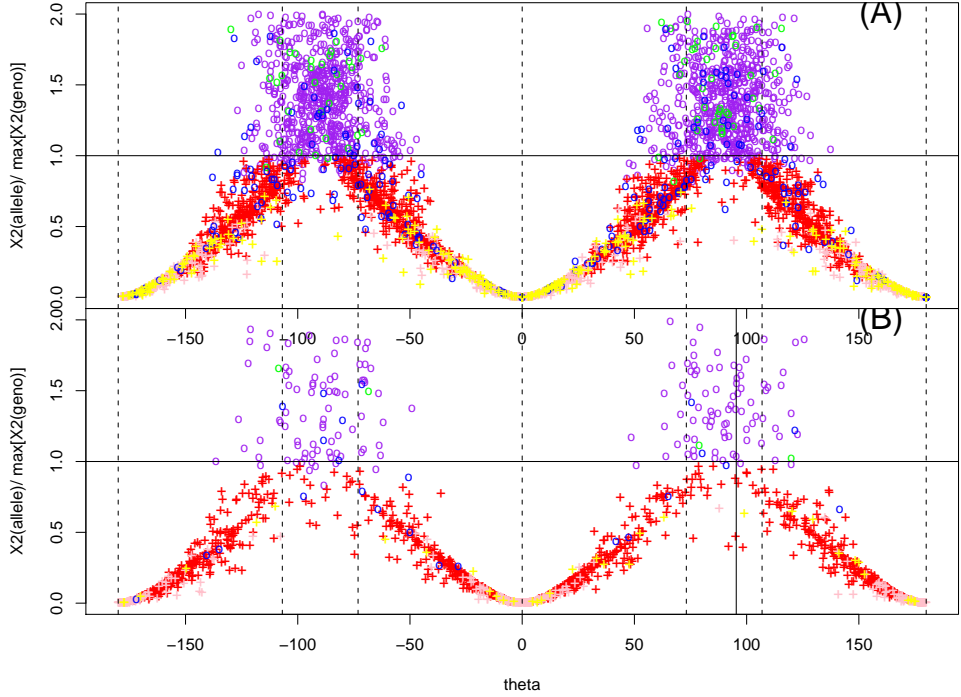}}
  \end{turn}
\caption{
The $X^2$ ratio $\lambda=X^2_{\rm allele}/\max(X^2_{\rm rec}, X^2_{\rm dom})$
as a function of the parameter $\theta = tan^{-1}(\delta/\delta_\epsilon)$.
The same color code for the 
six phases as used in Fig.\ref{fig:1} is also used here. 
For phases that favor the genotype-based test, $\lambda < 1$;
for those favoring allele-based test, $\lambda > 1$. (A) 100
samples per group with 5000 replicates; (B) 1000 samples
per group with 5000 replicates.
The vertical dashed lines correspond to angles of
$\pm$ 73.125$^\circ$ and $\pm$ 106.875$^\circ$, and the solid vertical
line corresponds to angle of 95.37$^circ$. 
\label{fig:2}}
\end{figure}

If we focus on the two major categories (phases I,II,III versus phases
IV,V,VI), we notice that the phase boundaries are radiuses. The
observation led to the following phase diagram by using a
single parameter $\theta = tan^{-1} (y/x) = tan^{-1} (\delta/\delta_\epsilon)$,
i.e., the angle between a radius and the x-axis. To measure
the relative advantage (disadvantage) of allele-based test
over genotype-based test,  we use the ratio of two $X^2$'s:
$\lambda= X^2_{\rm allele}/\max(X^2_{\rm rec}, X^2_{\rm dom})$.
Fig.\ref{fig:2} shows $\lambda$ as a function of $\theta$, 
using the simulation result in Fig.\ref{fig:1} (100 samples
per group and 1000 samples per group) and the same color code 
for six phases. 

Fig.\ref{fig:2} shows that within the range of
$ - 13 \pi/16 <  \theta < 13 \pi/16$ ($-73.125^\circ < \theta < 73.125^\circ$,
or $-3.2966 < \delta/\delta_\epsilon < 3.2966$),
the genotype-based test is favored over the allele-based test. 
Overlap of phases still occurs in Fig.\ref{fig:2}, 
indicating the phase separation is not perfect.
The allele-based test is much better than the 
genotype-based test when $\theta = \pi/2$ (90$^\circ$).
and the genotype-based test is much better than 
the allele-based test when $\theta = 0$ (or $\delta=0$).

The sample size per group does not affect the phase boundary between the
two major categories, though it does affect phases
within a major category. This observation can be understood
theoretically by the formula of $X^2$'s in Eq.(\ref{eq:x2-epsilon-2}):
the relative magnitude between $X^2_{\rm allele}$ and
$X^2_{\rm dom, HWD}$ or $X^2_{\rm rec, HWD}$ is independent
of $N$ as it is canceled out.

\begin{table}[bt]
\tbl{Count tables of marker genotype for a SNP within the gene PTPN22
\label{table:3} }
{\begin{tabular}{@{}c|cccccc@{}}
\hline
{} & $TT$ & $TC$ & $CC$ & total & $p_T$ & $\epsilon$  \\[1ex]
\hline
case & 16 &  245  &  677 & 938 & .147655
 & -0.004744  \\[1ex] 
control & 12  & 221 & 1168 & 1401  & .087438 
 & +0.000920   \\[1ex]
\hline
difference & & & &  & .060217 & -.005664 
\end{tabular}
}
\end{table}

\section{Illustration by a real dataset}

The genotype counts of a missense SNP in gene PTPN22
in Rheumatoid Arthritis samples and in control samples
are listed in Table \ref{table:3} (combining the
``discovery" dataset and the ``single sib" option in
the ``replication" dataset in
Ref. \refcite{begovich}). Our formula predicts
that $\theta = tan^{-1}(0.147655-0.087438)/(-0.004744-0.000920)
= tan^{-1} (-0.060217/0.005664) = 95.37^\circ$. This $\theta$
line is marked both in Fig.\ref{fig:1} and Fig.\ref{fig:2}
in solid lines,
and is within the phase where the allele-based test is preferred. 
Our calculation predicts
that the allele-based test and genotype-based test should
lead to similar result.\footnote{One difference however is
that the theoretical calculation is based on equal number
of samples in case and control group. In our example,
the sample size in two groups is slightly different.}
Indeed, $X^2_{\rm allele}$=41.10, $X^2_{\rm dom, HWD}$=42.26,
$X^2_{\rm rec, HWD}$=3.43, and allele-based and 
genotype-based test statistics are essentially the same.

\section{Hardy-Weinberg disequilibrium in the patient population
given a disease model}

In the population of patients (case group), a SNP marker
within the disease gene or in linkage disequilibrium with
the disease usually violates the Hardy-Weinberg equilibrium.
This fact has been used in the proposal of using  HWD in
case samples to map the disease gene\cite{feder}. The HWD
coefficient in the case group can be calculated if the
disease model is given\cite{wlee}, which is reproduced here.
Assuming the penetrance for $AA, Aa, aa$ genotypes to be
$f_{\rm AA}, f_{\rm Aa}, f_{\rm aa}$, the disease prevalence is
$K= f_{\rm AA} p_1^2 + f_{\rm Aa} 2p_1q_1 + f_{\rm aa} q_1^2$, 
and the genotype frequencies for the case group are 
(using the Bayes' theorem):
\begin{equation}
p_{\rm AA,aff}=\frac{f_{\rm AA} p_1^2}{K},
\hspace{0.2in}
p_{\rm Aa,aff}=\frac{f_{\rm Aa} 2p_1q_1}{K},
\hspace{0.2in}
p_{\rm aa,aff}=\frac{f_{\rm aa} q_1^2}{K}.
\end{equation}
The HWD coefficient for the case group is then\cite{wlee}:
\begin{equation}
\epsilon_1 = p_{\rm AA,aff} \cdot p_{\rm aa,aff} - \frac{ p_{\rm Aa,aff}^2}{4}
= \left( \frac{p_1q_1}{K} \right)^2 (f_{\rm AA} f_{\rm aa} -f_{\rm Aa}^2),
\end{equation}
and the HWD coefficient for the control group is assumed
to be zero ($\epsilon_2=0$).

If the disease model is multiplicative, i.e., $f_{\rm AA}/f_{\rm Aa}= f_{\rm Aa}/f_{\rm aa}$,
there is no HWD in the case group, so HWD can not be used to map
the disease gene. With $\delta_\epsilon=0-0=0$,  from 
the result in Sec. \ref{sec:2}, the allele-based test
is favored over the genotype-based test.  For  
dominant models, $f_{\rm AA} \approx f_{\rm Aa} = F$, and
$\epsilon_1 \propto F(f_{\rm aa} -F)$. Since we usually assume
low phenocopy rate, i.e., $f_{\rm aa} \approx 0$, the HWDc $\epsilon_1
\propto -F^2$ is negative. If the mutant allele $A$ is enriched in case
samples ($\delta = p_1 - p_2 > 0$), with the $\delta_\epsilon < 0$
in dominant models, we conclude that genotype-based test
is favored over allele-based tests. For recessive models,
$f_{\rm Aa} \approx f_{\rm aa} \approx 0$, 
$\epsilon_1 \propto 0 $, so the allele-based test is better.
For additive models, $f_{\rm Aa}= f_{\rm aa}+ \Delta$,
$f_{\rm AA}= f_{\rm aa}+ 2\Delta$, where $\Delta$ is the
contribution to the penetrance by adding one copy of the
mutant allele. The $\delta_\epsilon$ is
equal to   $\epsilon_1 \propto (f_{\rm aa}+2\Delta) f_{\rm aa} - (f_{\rm aa}+\Delta)^2
= -\Delta^2 < 0$. Thus genotype-based test is favored for
additive disease models.

\section{Discussion and future works}

The main point of this article is that genotype-based
test may take advantage of certain Hardy-Weinberg disequilibrium in
case samples to overcome the advantage of larger
sample sizes in allele-based tests. Another advantage of
the genotype-based test is that it tests two models and
picks the best one. This multiple testing might be corrected
by multiplying the $p$-value by a factor of 2 (Bonferroni
corrections), which was not done in this article. 
Whether correcting multiple testing or 
not is always under debate\cite{perneger1,aickin,perneger2},
but its effect on our problem is probably to shift the
phase boundary slightly. 

The $X^2$ test statistic calculation in this article was
all carried out assuming equal number of samples in case
and control group. Changing this assumption to unequal
number of samples per group is not difficult, but its
effect on the conclusion has not been examined.

Here we are addressing the type-I error of the test,
the $p$-value, which is determined by the $X^2$ test
statistic. For type-II error under alternative hypothesis,
usually a non-central $\chi^2$ distribution could be used\cite{patnaik}.
However, other alternatives to non-central $\chi^2$
distribution to calculate type-II error and the power
have been proposed\cite{bukszar}.

\section*{Acknowledgments}

W.L. acknowledges the support from The Robert S. Boas Center 
for Genomics and Human Genetics at the Feinstein Institute for
Medical Research. {\bf Note added on March 2008: There was
an error in the November 2006 version (Proc. of 5th APBC, Imperial
College Press, pp.185-194 (2007)) which affected Fig.1 and
Fig.2. These two figures as well as the relevant sentences
in the text have been corrected.}


\begin{thebibliography}{0}

\bibitem{agresti} 
A. Agresti.
\newblock {\sl Categorical Data Analysis} (Wiley-Interscience, 2002).

\bibitem{aickin} 
M. Aickin.
\newblock Other method for adjustment of multiple testing exists.
\newblock {\rm BMJ}, 318:127, 1999.

\bibitem{begovich} 
A.B.  Begovich, et al.,
\newblock A missense single-nucleotide polymorphism in a gene encoding a
protein tyrosine phosphatase (PTPN22) is associated with rheumatoid arthritis.
\newblock {\em Am. J. Human Genet.}, 75:330-337, 2004.


\bibitem{brookes} 
A.J. Brookes.
\newblock The essence of SNPs.
\newblock {\em Gene}, 234:177-186, 1999.

\bibitem{bukszar} 
J. Bukszar, E.J. van den Oord.
\newblock Accurate and efficient power calculations for $2\times m$ 
tables in unmatched case-control designs.
\newblock {\em Stat. in Med.}, 25:2623-2646, 2006.

\bibitem{burton} 
P.R. Burton, M.D. Tobin, J.K. Happer.
\newblock Key concepts in genetic epidemiology.
\newblock {\em Lancet}, 366:941-951, 2005.

\bibitem{clayton05} 
H.J. Cordell, D.G. Clayton.
\newblock Genetic association studies.
\newblock {\em Lancet}, 366:1121-1131, 2005.

\bibitem{feder} 
J.N. Feder, et al.,
\newblock A novel MHC class I-like gene is mutated in patients with 
hereditary haemochromatosis.
\newblock {\em Nature Genet.}, 13:399-408, 1996.


\bibitem{lee} 
A.T. Lee, W. Li, A. Liew, C. Bombardier, M. Weisman, E.M. Massarotti,
J. Kent, F. Wolfe, A.B. Begovich, P.K. Gregersen.
\newblock
The PTPN22 R620W polymorphism associates with RF positive rheumatoid arthritis
in a dose-dependent manner but not with HLA-SE status.
\newblock {\em Genes and Immunity}, 6:129-133,  2005.

\bibitem{wlee} 
W.C. Lee.
\newblock Searching for disease-susceptibility loci by testing for 
Hardy-Weinberg disequilibrium in a gene bank of affected individuals.
\newblock {\em Am. J. Epidemiology}, 158:397-400, 2003.

\bibitem{wliweb} 
W. Li, edited,
\newblock Bibliography: linkage disequilibrium analysis
\newblock URL: {\sl http://www.nslij-genetics.org/ld/}

\bibitem{landau} 
E.M. Lifshitz, L.D. Landau.
\newblock {\sl Statistical Physics: Course of Theoretical Physics, Volume 5},
3rd edition (Butterworth-Heinemann, 1980).  


\bibitem{patnaik} 
P.B. Patnaik.
\newblock The non-central $\chi^2$- and $F$-Distribution and their applications.
\newblock {\em Biometrika}, 36:202-232, 1949.

\bibitem{perneger1} 
T.V. Perneger.
\newblock What's wrong with Bonferroni adjustments.
\newblock {\rm BMJ}, 316:1236-1238, 1998. 

\bibitem{perneger2} 
T.V. Perneger.
\newblock Adjusting for multiple testing in studies is less important than other concerns.
\newblock {\rm BMJ}, 318:1288, 1999.

\bibitem{risch96} 
N. Risch, K. Merikangas.
\newblock The future of genetic studies of complex human diseases.
\newblock {\em Science}, 273:1516-1517, 1996.

\bibitem{sasieni} 
P.D. Sasieni.
\newblock From genotypes to genes: doubling the sample size.
\newblock {\em Biometrics}, 53:1253-1261, 1997.

\bibitem{tokuhiro} 
S.  Tokuhiro,  et al.
\newblock
An intronic SNP in a RUNX1 binding site of SLC22A4, encoding
an organic cation transporter, is associated with rheumatoid arthritis.
\newblock {\rm Nature Genet.}, 35:341-348, 2003.


\bibitem{weir} 
B.S. Weir.
\newblock {\sl Genetic Data Analysis II} (Sinauer Associates, 1996).


\bibitem{yamada} 
R. Yamada,  et al.,
\newblock Association between a single-nucleotide polymorphism in the promoter
of the human interleukin-3 gene and rheumatoid arthritis in Japanese
patients, and maximum-likelihood estimation of combinatorial effect
that two genetic loci have on susceptibility to the disease.
\newblock {\em Am. J. Hum. Genet.}, 68:674-685, 2001.
 

\end{thebibliography}
\end{document}